\documentclass[11pt, draftcls, onecolumn]{IEEEtran}
\usepackage[english]{babel}
\usepackage{amsmath,amsthm}
\usepackage{amsfonts}
\usepackage{graphicx}
\usepackage{verbatim}
\usepackage{amssymb}
\usepackage{epstopdf}
\usepackage{hyperref}
\usepackage{bbm}
\usepackage{cite,bbm,graphicx,amsmath,amssymb,mathrsfs,epsf} \usepackage{multirow}
\usepackage{subfigure,cite,graphicx,epsfig,amsmath,amssymb,mathrsfs,epsf,graphics,color,enumerate}
\allowdisplaybreaks[4]
\newtheorem{theorem}{Theorem}

\newtheorem{lem}{Lemma}

\newtheorem{rem}{Remark}
\DeclareMathOperator*{\argmin}{arg\,min}

\def \E{\operatorname{E}}

\begin{document}
\title{The Wiretapped Diamond-Relay Channel}%
\author{\IEEEauthorblockN{Si-Hyeon Lee and Ashish Khisti}\thanks{S.-H. Lee and A. Khisti are with the Department of Electrical and Computer Engineering,
University of Toronto, Toronto, Canada (e-mail: sihyeon.lee@utoronto.ca; akhisti@comm.utoronto.ca).  }}
\maketitle

\begin{abstract}
In this paper, we study a diamond-relay channel where the source is connected to $M$ relays through orthogonal links and the relays transmit to the destination over a wireless multiple-access channel in the presence of an eavesdropper. The eavesdropper not only observes the relay transmissions through another multiple-access channel, but also observes a certain number of source-relay links. The legitimate terminals know neither the eavesdropper's channel state information nor the location of source-relay links revealed to the eavesdropper except the total number of such links.

For this \emph{wiretapped diamond-relay channel}, we establish the optimal secure degrees of freedom. In the achievability part, our proposed scheme uses the source-relay links to transmit a judiciously constructed combination of message symbols, artificial noise symbols as well as fictitious message symbols associated with secure network coding. The relays use a combination of beamforming and interference alignment in their transmission scheme.
For the converse part, we take a genie-aided approach assuming that the location of wiretapped links is known. 
\end{abstract}
\section{Introduction}
Cloud Radio-Access Network (C-RAN) is a promising architecture to meet the demand for higher data rates in next generation wireless networks. In these systems, the base-stations act as relays and are connected via high-speed backhaul links to a cloud network. Encoding and decoding operations happen centrally in the cloud. The study of fundamental information theoretic limits and optimal coding techniques for such systems is a fertile area of research. 

Motivated by C-RAN, we study a model where the source is connected to $M$ relay terminals using orthogonal links with a fixed capacity. The relays transmit to the destination  over a wireless  multiple-access channel. 
Such a setup is known as the diamond-relay network~\cite{ScheinGallager:00, schein_thesis,TraskovKramer:07,KangLiu:11,BidokhtiKramer:arxiv15}. We study this model in the presence of an eavesdropper who can eavesdrop the orthogonal links from the source to the relays as well as the wireless transmission from the relays. We require that a message be transmitted reliably to the legitimate receiver, while keeping it confidential from the eavesdropper. We adopt the information-theoretic notion of confidentiality, widely used in the literature on the wiretap  channel~\cite{Wyner:75,CsiszarKorner:78,CheongHellman:78,KhistiWornell:10,KhistiWornell:10_2,relay-3,LiangPoor:08, relay-4, relay-2,   TekinYener:08,TekinYener:08general,GoelNegi:08, perron_thesis,relay-5, relay-6,YassaeeArefGohari:14,relay-7,LeeKhisti:15}. We thus refer to our setup as the {\em{wiretapped diamond-relay channel}}.

While the secrecy capacity of both the scalar Gaussian wiretap channel and its multi-antenna extension  can be achieved using Gaussian codebooks \cite{CheongHellman:78,KhistiWornell:10,KhistiWornell:10_2}, the optimal schemes in multiuser channels are considerably more intricate. Very recently, Xie and Ulukus~\cite{XieUlukus:14} studied the secure degrees of freedom  (d.o.f.) for one-hop Gaussian multiuser channels such as the multiple-access channel and the interference channel. It turns out that interference alignment \cite{CadambeJafar:08,MotahariOveisGharanMaddahAliKhandani:14} plays a central role in achieving the optimal secure d.o.f. While coding schemes based on Gaussian codebooks can only achieve zero secure d.o.f., the schemes presented in~\cite{XieUlukus:14} involve transmitting a combination of information and jamming signals at each transmitter and judiciously precoding them such that the noise symbols align at the legitimate receiver(s), yet mask information symbols at the eavesdropper. This approach has been extended for the case with no eavesdropper's channel state information (CSI) at the legitimate parties  in~\cite{XieUlukus:13blind,MukherjeeXieUlukus:arxiv15}.  We note that a combination of jamming and interference alignment is also required in the confidential MIMO broadcast channel with delayed channel state information~\cite{yang2013secrecy} and the compound MIMO wiretap channel~\cite{khisti2013artificial}.

The diamond-relay channel that we consider is a two-hop network where the first hop consists of orthogonal links from the source to the relays, while the second hop is a multiple-access channel from the relays to the destination. Such a network is considerably different from one-hop networks as the channel inputs from the relays need not be mutually independent. The source  can transmit common/independent message, common noise, or any combination of those to the relays. In~\cite{LeeZhaoKhisti:arxiv15}, for the diamond-relay channel where an eavesdropper can  wiretap \emph{only} the multiple-access portion of the channel, it is shown that transmitting common noise over the source-relay links  facilitates more efficient jamming for both the case with full CSI and the case with no eavesdropper's CSI. For example, a key constituent scheme proposed in \cite{LeeZhaoKhisti:arxiv15} for the case with no eavesdropper's CSI is  computation for jamming (CoJ) where the source  transmits a function of information and common noise symbols to two relays in order that the common noise symbols introduced to jam information symbols at the eavesdropper can be  canceled at the destination through suitable precoding. When there are more than two relays, this scheme operates  in a time-sharing basis in a way that only two source-relay links are utilized in each sub-scheme.\footnote{Similarly, the other constituent scheme in \cite{LeeZhaoKhisti:arxiv15} for the case with no eavesdropper's CSI  operates in a time-sharing basis in a way that only a single source-relay link is active in each sub-scheme. } 


In the present paper, the optimal  secure d.o.f. of the wiretapped diamond-relay channel is established for the case with no eavesdropper's CSI. 
In our wiretapped diamond-relay channel, the eavesdropper not only observes the relay transmissions through another multiple-access channel, but also observes a certain number of source-relay links. The legitimate terminals do not know which subset of source-relay links are revealed to the eavesdropper, except for the total number of such links. 
For achievability,  we note that it is not straightforward to extend the proposed schemes in \cite{LeeZhaoKhisti:arxiv15} because their sub-schemes require \emph{asymmetric} link d.o.f. and hence the amount of wiretapped information varies over time according to the unknown location of wiretapped links.
Thus, we first develop symmetric version of  the proposed schemes in~\cite{LeeZhaoKhisti:arxiv15} and then combine them with a secure network coding (SNC)~\cite{CaiYeung:02,CaiYeung:11}-like scheme to account for non-secure source-relay links. While the conventional SNC has been developed for fully wired networks, we incorporate SNC by utilizing the nature of wireless networks in a way that the additional randomness introduced in the source-relay links due to SNC is canceled at the destination by beamforming over the wireless multiple-access channel. To that end, we judiciously choose the generator matrix for SNC. From a technical point of view, the secrecy analysis involves accounting for the observations from the source-relay links as well as the multiple-access wiretap channel and is considerably more involved. Furthermore, the secure d.o.f. is shown to be the same  even when the knowledge of the compromised links is available. Indeed our converse is established via this genie-aided approach.

The rest of this paper is organized as follows. In Section~\ref{sec:model}, we formally state our model of wiretapped diamond-relay channel. The main result on the secure d.o.f. is presented in Section~\ref{sec:main}.  For the achievability part, our proposed schemes are described at a high-level in Section~\ref{sec:ach} and are rigorously stated in Appendix \ref{appendix:achievability}. The converse part is proved in Section~\ref{sec:conv}. We conclude this paper in Section~\ref{sec:conclusion}.

The following notation is used throughout the paper. For two integers $i$ and $j$, $[i:j]$ denotes the set $\{i,i+1,\cdots, j\}$. For constants $x_1,\cdots, x_k$ and $S\subseteq [1:k]$, $x_S$  denotes the  vector $(x_j: j\in S)$. This notation is naturally extended for vectors and random variables. For a sequence $x(1),x(2),\cdots$ of constants indexed by time, $x^k$ for positive integer $k$ denotes the vector $(x(j): j \in [1:k])$. This notation is naturally generalized for vectors and random variables. 
$\lfloor\cdot\rfloor$ denotes the floor function. For positive real number $\delta$ and positive integer $Q$, $\mathcal{C}(\delta, Q)$ denotes the PAM constellation $\delta\{- Q, - Q+1,\cdots, 0, \cdots, Q-1, Q\} $ of $(2Q+1)$ points with distance~$\delta$ between points. For positive integers $i$ and $j\in [0:i+1]$, $[j]_i$ denotes $j$ if $j\in [1:i]$, $i$ if $j=0$, and 1 if $j=i+1$. 

\section{Model}\label{sec:model}
Consider a diamond-relay channel that consists of an orthogonal broadcast channel from a source to $M\geq 2$ relays and a Gaussian multiple-access channel from the $M$ relays to a destination.  In the broadcast part, the source is connected to each relay through an orthogonal link of capacity $C$. In the multiple-access part, the channel output $Y_1(t)$ at time $t$ is given as 
\begin{align}
Y_1(t) = \sum_{k = 1}^M h_k(t) X_k(t) + Z_1(t), \label{eqn:y1_M}
\end{align}
in which $X_k(t)$ is the channel input at relay $k$, $h_k(t)$'s are channel fading coefficients, and $Z_1(t)$ is additive Gaussian noise with zero mean and unit variance. The transmit power constraint at relay $k$ is given as $\frac{1}{n}\sum_{t=1}^nX_{k}^2(t)\leq P$, where $n$ denotes the number of channel uses. 

\begin{figure}
  \centering
    \includegraphics[width=90mm]{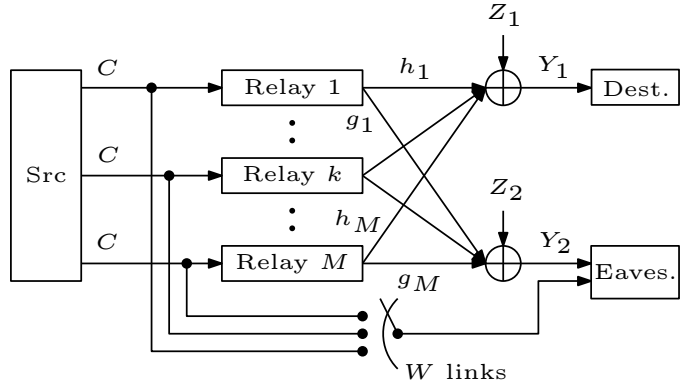}\\
  \caption{The wiretapped diamond-relay channel. }\label{fig:model}
\end{figure}

In this paper, we consider a scenario illustrated in Fig. \ref{fig:model}, where an eavesdropper wiretaps both the broadcast part and the multiple-access part of the diamond-relay channel. In the broadcast part, the eavesdropper can wiretap $W$ source-relay links. Let $N=M-W$ denote the number of secure source-relay links. We assume that the location of wiretapped links is unknown to the source, relays, and destination. In the multiple-access part, the eavesdropper observes $Y_2(t)$ at time $t$ given as 
\begin{align}
Y_2(t) = \sum_{k = 1}^M g_k(t) X_k(t) + Z_2(t), \label{eqn:y2_M}
\end{align}
where $g_k(t)$'s are channel fading coefficients and $Z_2(t)$ is additive Gaussian noise with zero mean and unit variance.

We assume a fast fading scenario where $h_k(t)$'s and $g_k(t)$'s are drawn in an i.i.d. fashion over time according to an arbitrary real-valued joint density function $f(h_1, \cdots, h_M, g_1, \cdots, g_M)$ satisfying that (i) all joint and conditional density functions are bounded and (ii) there exists a positive finite number $B$ such that 
$\frac{1}{B}\leq |h_k(t)|, |g_k(t)| \leq B$ for all $k\in [1:M]$.\footnote{The former condition implies that the probability that the channel fading coefficients are in a space of Lebesque measure zero is zero. The latter condition is a mild technical condition to avoid degenerate situations and has a vanishing impact on the d.o.f. because by choosing $B$ large enough, the omitted support set can be reduced to a negligible probability.}  For notational convenience, let  $\mathbf{h}(t)=(h_1(t)~\cdots ~h_M(t))$ and  $\mathbf{g}(t)=(g_1(t)~\cdots ~g_M(t))$ denote the legitimate channel state information (CSI) and the eavesdropper's CSI at time $t$, respectively.
We assume that the source does not know both the legitimate CSI and the eavesdropper's CSI and the eavesdropper knows both the CSI's.  The relays and destination are assumed to know  \emph{only} the legitimate CSI.\footnote{Although the relays are assumed to know the \emph{global} legitimate CSI, i.e., each relay knows all the channel fading coefficients to the destination, our proposed schemes require relays to know only some \emph{local} legitimate CSI. } 

 A $(2^{nR}, n)$  code consists of a message $G\sim \mbox{Unif}[1:2^{nR}]$,\footnote{$\mbox{Unif}[S]$ for a set $S$ denotes the uniform distribution over $S$. When $S=[i:j]$, we use $\mbox{Unif}[i:j]$ instead of $\mbox{Unif}[[i:j]]$.} a stochastic encoder at the source that (randomly) maps $G\in [1:2^{nR}]$ to $(S_1^n, \cdots, S_M^n)\in \mathcal{S}_1^n\times \cdots \times \mathcal{S}_M^n$ such that $\frac{1}{n}H(S_k^n)\leq C$ for $k\in [1:M]$, a stochastic encoder at time $t\in[1:n]$ at relay $k\in [1:M]$ that (randomly) maps $(S_k^n, X_k^{t-1}, \mathbf{h}^t)$ to $X_k(t) \in \mathcal{X}_k$, and a decoding function at the destination that (randomly) maps $(Y_1^n, \mathbf{h}^n)$  to $\hat{G} \in [1:2^{nR}]$. The probability of error is given as  $P_e^{(n)}=P(\hat{G}\neq G)$. A secrecy rate of $R$ is said to be achievable if there exists a sequence of $(2^{nR},n)$ codes such that $\lim_{n\rightarrow \infty}P_e^{(n)}=0$  and 
\begin{align}
\lim_{n\rightarrow \infty}\frac{1}{n}I(W;S_{T}^n,Y_2^n|\mathbf{h}^n, \mathbf{g}^n)=0\label{eqn:sec_cond}
\end{align}
for all $T\subseteq [1:M]$ such that $|T|=W$. The secrecy capacity is the supremum of all achievable secrecy rates.

We also consider the case where the location of wiretapped links is known to all parties. Let $T$ denote the index set of relays whose links from the source are wiretapped. In this case,  an achievable secrecy rate is defined by requiring \eqref{eqn:sec_cond}  to be satisfied for the known~$T$.

In this paper, we are interested in asymptotic behavior of the secrecy capacity when $P$ tends to infinity. We say a d.o.f. tuple $(\alpha, d_s)$ is achievable if a secrecy rate $R$ such that $\lim_{P\rightarrow \infty} \frac{R}{\frac{1}{2}\log P}=d_s$ is achievable when $\lim_{P\rightarrow \infty} \frac{C}{\frac{1}{2}\log P}=\alpha$. A secure d.o.f. $d_s(\alpha)$ is the maximum $d_s$ such that $(\alpha, d_s)$ is achievable.  According to the context, $d_s$ denotes $d_s(\alpha)$.

\begin{rem}
Our model guarantees secrecy from a certain level of relay collusion, i.e., from collusion of any set of up to $W$ relays when the location of wiretapped links is unknown and from collusion of any set of relays that have wiretapped links when it is known. 
\end{rem}
\begin{rem}
Our achievability results hold for constant channel gains as well as slow-fading channels. Furthermore, our achievability results can be generalized for complex channel fading coefficients by applying \cite[Lemma 7]{MaddahAli:09} in our analysis of interference alignment.
\end{rem}


\section{Main Result} \label{sec:main}
In this section, we state the main result of this paper. 
Let us first present the secure d.o.f. of the wiretapped diamond-relay channel when $N=M$, which was established in \cite[Theorem 6]{LeeZhaoKhisti:arxiv15}.
\begin{theorem}\label{thm:same}
The secure d.o.f. of the wiretapped diamond-relay channel when $N=M$, i.e., the eavesdropper does not wiretap the broadcast part, is equal to 
\begin{align}
d_{s} &= \min\left\{M\alpha,\frac{M\alpha + M - 1}{M + 1},1\right\}.
\end{align}
\end{theorem} 
For the achievability part of Theorem~\ref{thm:same}, two constituent schemes were proposed in \cite{LeeZhaoKhisti:arxiv15}. One scheme  is a time-shared version of the blind cooperative jamming (BCJ)  scheme, where in each sub-scheme, a single relay operates as a source and the other relays operate as helpers according to the BCJ scheme \cite{XieUlukus:13blind}. The other scheme  is a time-shared version of the CoJ scheme, where in each sub-scheme, the source and a pair of relays operate according to the CoJ scheme \cite[Scheme 5]{LeeZhaoKhisti:arxiv15} while the other relays remain idle. For the converse part of Theorem~\ref{thm:same}, a technique was introduced in \cite{LeeZhaoKhisti:arxiv15} that captures the trade-off between the message rate and the amount of individual randomness injected at each relay.

\begin{figure}
  \centering
    \includegraphics[width=120mm]{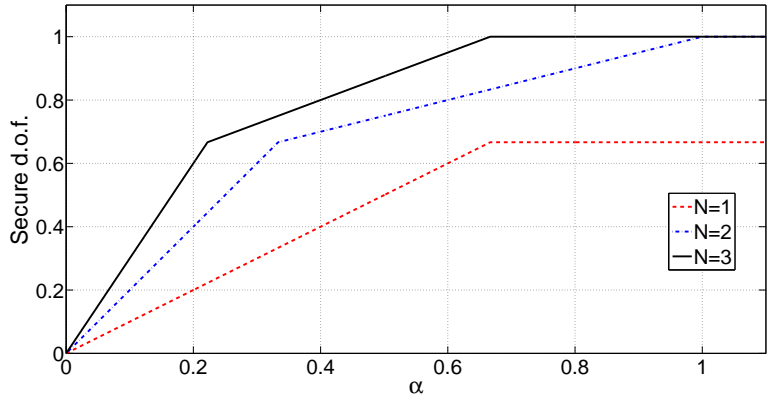}\\
  \caption{Secure d.o.f. of the wiretapped diamond-relay channel when $M=3$ and $N=1,2,3$. }\label{fig:thm2}
\end{figure}

The following theorem  establishes the secure d.o.f. of the wiretapped diamond-relay channel for the general case.
\begin{theorem} \label{thm:less}
The secure d.o.f. of the wiretapped diamond-relay channel is equal to 
\begin{align}
d_{s} &= \begin{cases}
\min\left\{\alpha, \frac{M-1}{M}\right\},  &N=1\\
\min\left\{N\alpha,\frac{N\alpha + M - 1}{M + 1},1\right\}, &N\geq 2
\end{cases}
\end{align}
for both the cases with and without the knowledge of location of wiretapped links. 
\end{theorem} 
Theorem~\ref{thm:less} indicates that the secure d.o.f. is the same for the cases with and without the knowledge of location of wiretapped links. According to Theorem~\ref{thm:less}, the secure d.o.f. is at most $\frac{M-1}{M}$ when the number of secure links is one. When the number of secure links is one and its location is known, a natural strategy would be to send the message over the secure link and send nothing over the wiretapped links. Then, the multiple-access part becomes the wiretap channel with $M-1$ helpers whose secure d.o.f. is shown to be $\frac{M-1}{M}$ in \cite{XieUlukus:13blind}. When $N\geq 2$, we can interpret Theorem~\ref{thm:less} in a way that the secure d.o.f. is decreased as if the link d.o.f. $\alpha$ was decreased by a factor of $\frac{N}{M}$. Because only $N$ out of $M$ links are secure, it is intuitive that the information that can be securely sent over the broadcast part is decreased by a factor of $\frac{N}{M}$. In Fig. \ref{fig:thm2}, the secure d.o.f. of the wiretapped diamond-relay channel is illustrated when $M=3$ and $N=1,2,3$.

The achievability part of Theorem~\ref{thm:less} when the location of wiretapped links is known can be easily proved by generalizing the proposed  schemes in \cite{LeeZhaoKhisti:arxiv15}. However, when  the location of wiretapped links is unknown, it is not straightforward to extend the proposed schemes in \cite{LeeZhaoKhisti:arxiv15} because their sub-schemes require \emph{asymmetric} link d.o.f. and hence the amount of wiretapped information depends on the unknown location of wiretapped links. To resolve this issue, we first propose \emph{simultaneous} BCJ (S-BCJ) and \emph{simultaneous} CoJ (S-CoJ) schemes for the case with the knowledge of location of wiretapped links. Then we incorporate SNC on top of those schemes for the case without the knowledge of location of wiretapped links. Our proposed schemes are described at a high-level in Section~\ref{sec:ach} and rigorously stated in  Appendix \ref{appendix:achievability}. The converse part of Theorem~\ref{thm:less} is proved in Section~\ref{sec:conv}.

\section{Achievability} \label{sec:ach}
To prove the achievability part of Theorem~\ref{thm:less}, it suffices to show the achievability of the following corner points: $(\alpha,d_s)=(\frac{M-1}{MN}, \frac{M-1}{M})$ for $N\geq 1$ and $(\alpha,d_s)=(\frac{2}{N}, 1)$ for $N\geq 2$.\footnote{Note that for $N=1$,  $d_s=\min\left\{\alpha, \frac{M-1}{M}\right\}$  can be shown to be achievable by time-sharing between $(\alpha,d_s)=(0,0)$ and $(\alpha,d_s)=(\frac{M-1}{M}, \frac{M-1}{M})$. For $N\geq 2$, $d_s=\min\left\{N\alpha,\frac{N\alpha + M - 1}{M + 1},1\right\}$ can be shown to be achievable by time-sharing among $(\alpha,d_s)=(0,0)$, $(\alpha,d_s)=(\frac{M-1}{MN}, \frac{M-1}{M})$, and $(\alpha,d_s)=(\frac{2}{N}, 1)$.} 
  
In the following, we first propose S-BCJ scheme achieving  $(\alpha,d_s)=(\frac{M-1}{M^2}, \frac{M-1}{M})$  and S-CoJ scheme achieving $(\alpha,d_s)=(\frac{2}{M}, 1)$ for the special case of $N=M$. Then, we generalize them for the general case of $N\leq M$ with the knowledge of location of wiretapped links. Subsequently, we incorporate SNC on top of the S-BCJ and S-CoJ schemes for the case of $N\leq M$ without the knowledge of location of wiretapped links. To provide main intuition behind our schemes, let us give a high-level description of our schemes in this section. A  detailed description with rigorous analysis is relegated to Appendix \ref{appendix:achievability}. 

\begin{figure*}[t]
 \centering
   \subfigure[]
  {  \includegraphics[width=160mm]{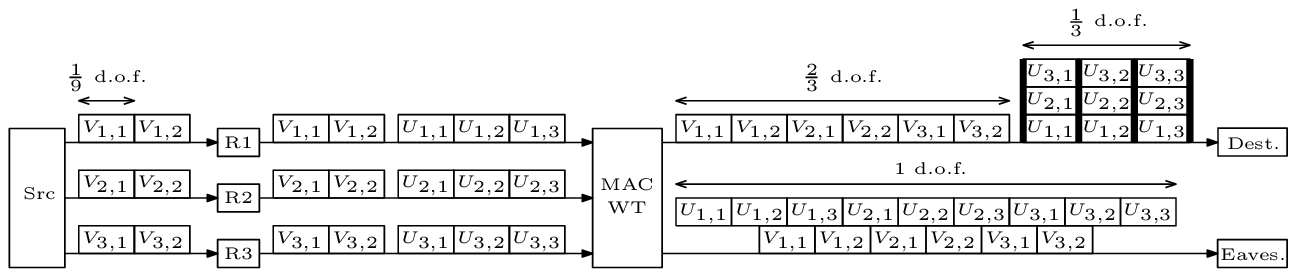}}
     \subfigure[]
  {  \includegraphics[width=130mm]{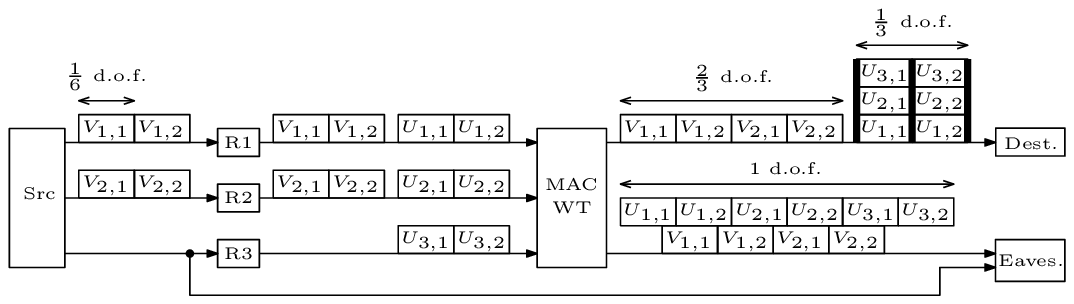}}
  \subfigure[]
  {  \includegraphics[width=135mm]{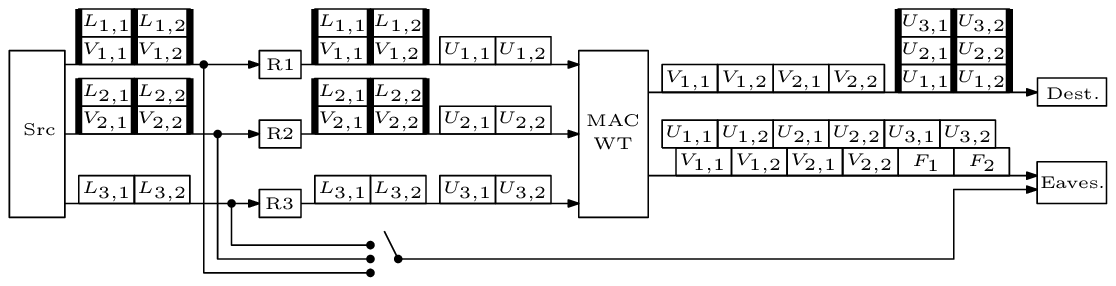}}
   \caption{(a): S-BCJ scheme for $M=N=3$, (b): S-BCJ scheme for $M=3$ and $N=2$ when the last link is known to be wiretapped, (c): S-BCJ-SNC scheme for $M=3$ and $N=2$ when the location of wiretapped link is unknown. Rectangles labeled with $V$, $U$, $F$ represent message, noise, and fictitious message symbols, respectively, and a rectangle labeled with $L$ represents a linear combination of fictitious message symbols. Each symbol (each rectangular) has $\frac{1}{9}$ d.o.f. in (a) and has $\frac{1}{6}$ d.o.f. in (b) and (c).  A column of symbols with fat side lines implies that those symbols are aligned and occupy the d.o.f. of a single symbol. } \label{fig:SBCJ}
  \end{figure*}
  \begin{figure}[t]
 \centering
   \subfigure[]
  {   \includegraphics[width=85mm]{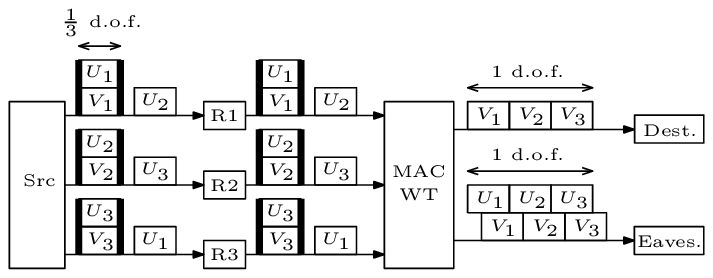}}
     \subfigure[]
  {   \includegraphics[width=82mm]{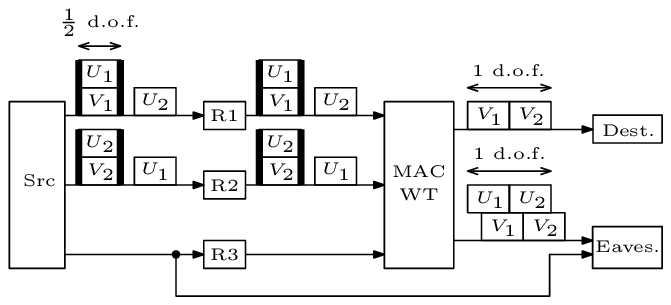}}
  \subfigure[]
  {    \includegraphics[width=87mm]{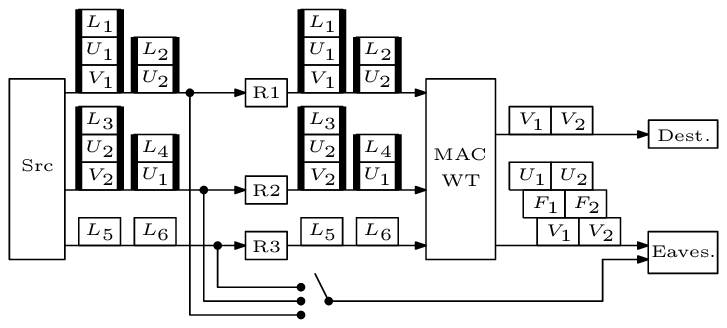}}
   \caption{(a): S-CoJ scheme for $M=N=3$, (b): S-CoJ scheme for $M=3$ and $N=2$ when the last link is known to be wiretapped, (c): S-CoJ-SNC scheme for $M=3$ and $N=2$ when the location of wiretapped link is unknown. Rectangles labeled with $V$, $U$, $F$ represent message, noise, and fictitious message symbols, respectively, and a rectangle labeled with $L$ represents a linear combination of fictitious message symbols. Each symbol (each rectangular) has $\frac{1}{3}$ d.o.f. in (a) and has $\frac{1}{2}$ d.o.f. in (b) and (c).  A column of symbols with fat side lines implies that those symbols are aligned and occupy the d.o.f. of a single symbol. } \label{fig:SCOJ}
  \end{figure}
  
In Fig. \ref{fig:SBCJ} and Fig. \ref{fig:SCOJ}, our proposed schemes are illustrated for $M=3$, where rectangles labeled with $V$, $U$, $F$ represent message, noise, and fictitious message symbols, respectively, and a rectangle labeled with $L$ represents a linear combination of fictitious message symbols.  A column of symbols with fat side lines implies that those symbols are aligned and occupy the d.o.f. of a single symbol.

\subsection{Special case of $N=M$}
\subsubsection{S-BCJ scheme for $N=M$ achieving $(\alpha, d_s)=(\frac{M-1}{M^2}, \frac{M-1}{M})$} The source represents the message of d.o.f. $\frac{M-1}{M}$ as a vector $V=(V_{k,j}: k\in [1:M], j\in [1:M-1])$ of independent message symbols each of d.o.f. $\frac{1}{M^2}$ and sends $V_k=(V_{k,j}:j\in [1:M-1])$ to relay $k$, which requires $\alpha=\frac{M-1}{M^2}$. Then, relay $k$ sends $V_k$ together with  $M$ independent noise symbols $(U_{k,j}: j\in [1:M])$ each of d.o.f. $\frac{1}{M^2}$ in a way that (i) for each $j\in [1:M]$, $U_{k,j}$'s for $k\in [1:M]$ are aligned at the  destination and (ii) $V_{k,j}$'s can be distinguished by the destination. Since  $U_{k,j}$'s are not aligned and occupy a total of 1 d.o.f at the eavesdropper, the message can be shown to be secure.

\subsubsection{S-CoJ scheme for $N=M$ achieving $(\alpha, d_s)=(\frac{2}{M}, 1)$}  The source represents the message of d.o.f. 1 as a vector $(V_1, \cdots, V_M)$ of independent message symbols and generates $M$ independent noise symbols $(U_1, \cdots, U_M)$, where each $V_k$ and $U_k$ has a d.o.f. $\frac{1}{M}$. The source sends $(V_k+U_k, U_{[k+1]_ M})$ to relay $k$, which requires $\alpha=\frac{2}{M}$. Then, the relays send what they have received in a way that (i) each of $U_k$'s is beam-formed in the null space of the destination's channel and (ii) $V_k$'s can be distinguished by the destination. Since $U_k$'s are not aligned and occupy a total of 1 d.o.f at the eavesdropper, the message can be shown to be secure.

\subsection{General case of $N\leq M$ with the knowledge of location of wiretapped links}\label{subsec:less_w}
In this case, we send nothing over the wiretapped links. Without loss of generality, let us assume that the first $N$ links are secure. 
\subsubsection{S-BCJ scheme  achieving $(\alpha, d_s)=(\frac{M-1}{MN}, \frac{M-1}{M})$}
The source represents the message of d.o.f. $\frac{M-1}{M}$ as a vector $V=(V_{k,j}: k\in [1:N], j\in [1:M-1])$ of independent message symbols each of d.o.f. $\frac{1}{MN}$ and sends $V_k=(V_{k,j}:j\in [1:M-1])$ to relay $k\in [1:N]$ and sends nothing to relay $i\in [N+1:M]$, which requires $\alpha=\frac{M-1}{MN}$. Then, relay $k\in [1:M]$ sends what it has received together with  $N$ independent noise symbols $(U_{k,j}: j\in [1:N])$ each of d.o.f. $\frac{1}{MN}$ in a way that (i) for each $j\in [1:N]$, $U_{k,j}$'s for $k\in [1:M]$ are aligned at the destination and (ii) $V_{k,j}$'s can be distinguished by the destination. Since $U_{k,j}$'s are not aligned and occupy a total of 1 d.o.f at the eavesdropper, the message can be shown to be secure.

\subsubsection{S-CoJ scheme  for $N\geq 2$ achieving $(\alpha, d_s)=(\frac{2}{N}, 1)$}  The source represents the message of d.o.f. 1 as a vector $(V_1, \cdots, V_N)$ of independent message symbols and generates $N$ independent noise symbols $(U_1, \cdots, U_N)$, where each $V_k$ and $U_k$ has a d.o.f. $\frac{1}{N}$. The source sends  $(V_k+U_k, U_{[k+1]_ N})$ to relay $k\in [1:N]$ and sends nothing to relay $i\in [N+1:M]$, which requires $\alpha=\frac{2}{N}$. Then, the relays send what they have received in a way that  (i) each of $U_k$'s is beam-formed in the null space of the destination's channel and (ii) $V_k$'s can be distinguished by the destination. Since $U_k$'s are not aligned and occupy a total of 1 d.o.f at the eavesdropper, the message can be shown to be secure.

\subsection{General case of $N\leq M$ without the knowledge of location of wiretapped links}
When the location of wiretapped links is unknown, the information sent over the source-relay links in the S-BCJ and S-CoJ schemes should be masked by additional randomness. To that end, we introduce independent fictitious message symbols whose total d.o.f. is the same as the total d.o.f. of wiretapped links. Then, in the S-BCJ and S-CoJ schemes assuming \emph{a certain location} of wiretapped links (e.g., the last $W$ links), we add a linear combination of these fictitious message symbols to each element sent from the source to the relays. By doing so, the message can be shown to be secure \emph{regardless of the location} of wiretapped links, while the fictitious message symbols are beam-formed in the null space of the destination and hence the achievable secure d.o.f. is not affected. 

This technique of masking the information by adding a linear combination of fictitious message symbols is similar to SNC~\cite{CaiYeung:02,CaiYeung:11}, and hence we call our schemes S-BCJ-SNC and S-CoJ-SNC.  However, it should be noted that our schemes assume computation over \emph{real numbers} to enable the beam-forming of fictitious message symbols over the multiple-access part,  while the conventional SNC has been developed for fully wired networks and assumes computation over \emph{finite field}.

\subsubsection{S-BCJ-SNC scheme achieving $(\alpha, d_s)=(\frac{M-1}{MN}, \frac{M-1}{M})$}
As in the S-BCJ scheme when the last $W$ links are wiretapped, the source represents the message of d.o.f. $\frac{M-1}{M}$ as a vector $V=(V_{k,j}: k\in [1:N], j\in [1:M-1])$ of independent message symbols each of d.o.f. $\frac{1}{MN}$. To mask each symbol, we introduce a vector $F=(F_k: k\in [1:W(M-1)])$ of independent fictitious message symbols each of d.o.f. $\frac{1}{MN}$ and generate a vector $L=(L_{k,j}: k\in [1:M], j\in[1:M-1])$ of  linear combinations of fictitious message symbols by computing $L=F\Gamma$ for some integer matrix $\Gamma$. Now, the source sends $(V_{k,j}+L_{k,j}:j\in [1:M-1])$ to relay $k\in [1:N]$ and sends $(L_{i,j}:j\in [1:M-1])$ to relay $i\in [N+1:M]$, which requires $\alpha=\frac{M-1}{MN}$. Then, relay $k\in [1:M]$ sends what it has received together with  $N$ independent  noise symbols $(U_{k,j}: j\in [1:N])$ each of d.o.f. $\frac{1}{MN}$ in a way that (i) for each $j\in [1:N]$, $U_{k,j}$'s for $k\in [1:M]$ are aligned at the  destination, (ii) each of $F_k$'s is beam-formed in the null space of the destination's channel, and (iii) $V_{k,j}$'s can be distinguished by the destination. By judiciously choosing the generator matrix $\Gamma$, it can be shown that the information leakage is zero (in the d.o.f. sense). 
 
\subsubsection{S-CoJ-SNC scheme for $N\geq 2$ achieving $(\alpha, d_s)=(\frac{2}{N}, 1)$}
As in the S-CoJ scheme when the last $W$ links are wiretapped, the source represents the message of d.o.f. 1 as a vector $(V_1, \cdots, V_N)$ of independent message symbols and generates $N$ independent noise symbols $(U_1, \cdots, U_N)$, where each $V_k$ and $U_k$ has a d.o.f. $\frac{1}{N}$. Then, for a vector $F=(F_k: k\in [1:2W])$ of independent fictitious message symbols each of d.o.f. $\frac{1}{N}$, a vector $L=(L_{k}: k\in [1:2M])$ of linear combinations of fictitious message symbols is generated by computing $L=F\Gamma$ for some integer matrix $\Gamma$. 
Now, the source sends $(V_k+U_k+L_{2k-1}, U_{[k+1]_ N}+L_{2k})$  to relay $k\in [1:N]$ and sends $(L_{2i-1}, L_{2i})$ to relay $i\in [N+1:M]$, which requires $\alpha=\frac{2}{N}$. Then, the relays send what they have received in a way that (i) each of $U_k$'s and $F_k$'s is beam-formed in the null space of the destination's channel and (ii) $V_{k}$'s can be distinguished by the destination. By judiciously choosing the generator matrix $\Gamma$, it can be shown that the information leakage is zero (in the d.o.f. sense). 

\section{Converse} \label{sec:conv}
It suffices to prove the converse part of Theorem~\ref{thm:less} for the case with the knowledge of location of wiretapped links.
\subsection{Proof of the converse part of Theorem~\ref{thm:less} for the case with the knowledge of location of wiretapped links}
For the Gaussian multiple-access wiretap channel, it is shown in \cite[Section 4.2.1]{MukherjeeXieUlukus:arxiv15} that there is no loss of secure d.o.f. if we consider the following deterministic model with integer-input and integer-output, instead of  \eqref{eqn:y1_M} and \eqref{eqn:y2_M}: 
\begin{align}
Y_1(t)=\sum_{k=1}^M \lfloor h_k(t)X_k(t)\rfloor,~ Y_2(t)=\sum_{k=1}^M \lfloor g_k(t)X_k(t)\rfloor \label{eqn:ch_M}
\end{align}
with the constraint 
\begin{align}
X_k(t)\in \{0,1,\ldots,\lfloor \sqrt{P}\rfloor\}, k=1,\ldots, M. \label{eqn:pw_M}
\end{align}
Likewise, it can be shown that there is no loss of secure d.o.f. in considering the deterministic model
 \eqref{eqn:ch_M} and \eqref{eqn:pw_M} for the multiple-access part of our model.\footnote{We omit a formal proof as it is straightforward from that in \cite[Section 4.2.1]{MukherjeeXieUlukus:arxiv15}.} Hence, in this section, let us assume that the multiple-access part is given as \eqref{eqn:ch_M} and \eqref{eqn:pw_M}, instead of  \eqref{eqn:y1_M} and \eqref{eqn:y2_M}. 

Without loss of generality, we assume that the first $N$ links are secure, i.e., $T=[N+1:M]$. Furthermore, we assume that $\mathbf{g}^n$ in addition to $\mathbf{h}^n$ is available at the destination, which only possibly increases the secure d.o.f. Hence,  $\mathbf{h}^n$ and $\mathbf{g}^n$  are conditioned in every entropy and mutual information terms in this section, but are omitted for notational convenience. In the following, $c_i$'s for $i=1,2,3,\ldots$ are used to denote positive constants that do not depend on $n$ and $P$.

First, we obtain 
\begin{align}
nR &\overset{(a)}\leq I(W;Y_1^n,S^n_{[N+1:M]}) + nc_1\\
&\overset{(b)}{\leq} I(W;Y_1^n,S^n_{[N+1:M]}) - I(W;Y_2^n,S^n_{[N+1:M]}) + nc_2 \label{eqn:public_fano}\\
& \leq I(W;Y_1^n|Y_2^n,S^n_{[N+1:M]}) + nc_2 \label{eqn:one_dof}\\
& \leq H(Y_1^n|Y_2^n,S^n_{[N+1:M]})+ nc_2\label{eqn:public_1}\\
& = H(Y_1^n,Y_2^n|S^n_{[N+1:M]}) - H(Y_2^n|S^n_{[N+1:M]}) + nc_2 \\
& \leq H(X_{[1:M]}^n,Y_1^n,Y_2^n|S^n_{[N+1:M]}) - H(Y_2^n|S^n_{[N+1:M]}) + nc_2\\
&=H(X_{[1:M]}^n|S^n_{[N+1:M]})+ H(Y_1^n,Y_2^n|X_{[1:M]}^n,S^n_{[N+1:M]}) - H(Y_2^n|S^n_{[N+1:M]}) + nc_2\\
&\overset{(c)}{=} H(X_{[1:M]}^n|S^n_{[N+1:M]}) - H(Y_2^n|S^n_{[N+1:M]}) + nc_2\label{eqn:same}\\
&\leq H(X_{[1:M]}^n, S^n_{[1:N]}|S^n_{[N+1:M]]})  - H(Y_2^n|S^n_{[N+1:M]}) + nc_2\\
&\leq H(S^n_{[1:N]})+ H(X_{[1:M]}^n|S^n_{[1:M]}) - H(Y_2^n|S^n_{[N+1:M]}) + nc_2\\
&\leq nNC +\sum_{i=1}^MH(X_i^n|S^n_i)- H(Y_2^n|S^n_{[N+1:M]})+nc_2. \label{eqn:cov12}
\end{align}
where $(a)$ is due to the Fano's inequality, $(b)$ is from the secrecy constraint, and $(c)$ is because the deterministic model  \eqref{eqn:ch_M} is assumed. 

Next, to bound $H(X_{i}^n|S^n_{i})$ for $i\in [1:M]$, we start from \eqref{eqn:public_fano} to obtain 
\begin{align}
nR &\leq I(W;Y_1^n,S^n_{[N+1:M]}) - I(W;Y_2^n,S^n_{[N+1:M]}) + nc_2\\
&\leq  I(W;Y_1^n,S^n_{[N+1:M]}) - I(W; S^n_{[N+1:M]}) + nc_2 \\
&\leq I(W;Y_1^n|S^n_{[N+1:M]}) + nc_2 \\
&\leq I(S^n_{[1:N]};Y_1^n|S^n_{[N+1:M]})+ nc_2  \label{eqn:cov_bc}\\
&\leq H(Y_1^n|S^n_{[N+1:M]})-H(Y_1^n|S^n_{[1:M]})+ nc_2 \\
&\overset{(a)}{\leq} H(Y_1^n|S^n_{[N+1:M]})-H(X_i^n|S^n_i)+ nc_4,  \label{eqn:cov22}
\end{align}
where $(a)$ is due to the following chain of inequalites: 
\begin{align}
H(Y_1^n|S^n_{[1:M]})&= H\big(\big\{\sum_{i = 1}^M \lfloor h_i(t) X_i(t)\rfloor\big\}_{t=1}^n |S^n_{[1:M]}\big)\\
&\geq H\big(\big\{\sum_{i = 1}^M \lfloor h_i(t) X_i(t)\rfloor\big\}_{t=1}^n |S^n_{[1:M]},X_{i^c}^n\big)\\
&=  H\big(\big\{ \lfloor h_i(t) X_i(t)\rfloor\big\}_{t=1}^n |S^n_{[1:M]},X_{i^c}^n\big)\\
&=H( X_i^n, \big\{ \lfloor h_i(t) X_i(t)\rfloor\big\}_{t=1}^n  |S^n_{[1:M]}, X_{i^c}^n)-H(X_i^n|\big\{ \lfloor h_i(t) X_i(t)\rfloor\big\}_{t=1}^n ,S^n_{[1:M]}, X_{i^c}^n)\\
&=H( X_i^n  |S^n_{[1:M]}, X_{i^c}^n)-H(X_i^n|\big\{ \lfloor h_i(t) X_i(t)\rfloor\big\}_{t=1}^n ,S^n_{[1:M]}, X_{i^c}^n)\\
&\geq H( X_i^n  |S^n_{[1:M]}, X_{i^c}^n)-\sum_{t=1}^nH(X_i(t)|\lfloor h_i(t) X_i(t)\rfloor)\\
&\overset{(a)}{\geq} H( X_i^n|S^n_{[1:M]}, X_{i^c}^n)-nc_3\\
&\overset{(b)}{=} H( X_i^n|S^n_i)-nc_3,
\end{align}
where $i^c$ denotes $[1:M]\setminus \{i\}$, $(a)$ is from \cite[Lemma 2]{MukherjeeXieUlukus:arxiv15}, and $(b)$ is due to the Markov chain\footnote{We remind that $\mathbf{h}^n$ and $\mathbf{g}^n$ are conditioned in every entropy and mutual information terms in this section.} $X_i^n - (\mathbf{h}^n, \mathbf{g}^n, S^n_i) - (X_{i^c}^n, S^n_{i^c})$.

Now, by combining \eqref{eqn:cov12} and \eqref{eqn:cov22}, we have 
\begin{align}
(M+1)nR&\leq nNC+MH(Y_1^n|S^n_{[N+1:M]}) -H(Y_2^n|S^n_{[N+1:M]})+nc_{5} \\
&= nNC+(M-1)H(Y_1^n|S^n_{[N+1:M]})+H(Y_1^n|S^n_{[N+1:M]})-H(Y_2^n|S^n_{[N+1:M]})+nc_{5}. \label{eqn:diff_pre}
\end{align}
Furthermore, $H(Y_1^n|S^n_{[N+1:M]})-H(Y_2^n|S^n_{[N+1:M]})$ can be bounded as follows: 
\begin{align}
&H(Y_1^n|S^n_{[N+1:M]})-H(Y_2^n|S^n_{[N+1:M]})\cr
&\leq \max_{s^n_{[N+1:M]}}\Big\{H(Y_1^n|S^n_{[N+1:M]}=s^n_{[N+1:M]}) -H(Y_2^n|S^n_{[N+1:M]}=s^n_{[N+1:M]})\Big\} \\
&\leq \max_{p(x_1^n, \cdots, x_M^n) \in \mathcal{P}_{X_{[1:M]}^n} }\Big\{H(Y_1^n)-H(Y_2^n)\Big\}\\
&\overset{(a)}\leq n\cdot o(\log P) \label{eqn:diff}
\end{align}
where $\mathcal{P}_{X_{[1:M]}^n} $ denotes the set of all possible distributions of codewords satisfying the power constraint \eqref{eqn:pw_M} and $(a)$ is from \cite[Section 6]{DavoodiJafar:14}. 

By substituting \eqref{eqn:diff} to \eqref{eqn:diff_pre}, we obtain $d_s\leq \frac{N\alpha+M-1}{M+1}$. On the other hand, from \eqref{eqn:cov_bc}, we  have $d_s\leq \min(N\alpha, 1)$.
Therefore, 
\begin{align}
d_s\leq \min\left\{N\alpha, \frac{N\alpha+ M-1}{M+1}, 1\right\}.
\end{align}

For the special case of $N=1$, to derive a tighter bound, we start from \eqref{eqn:same}: 
\begin{align}
nR&\leq H(X_{[1:M]}^n|S^n_{[2:M]}) - H(Y_2^n|S^n_{[2:M]}) + nc_2\\
&\leq H(X_{1}^n|S^n_{[2:M]})+\sum_{i=2}^M H(X_{i}^n|S^n_{i}) - H(Y_2^n|S^n_{[2:M]}) + nc_2. \label{eqn:cov_1}
\end{align}

Now, by combining \eqref{eqn:cov_1} and \eqref{eqn:cov22}, we have
\begin{align}
MnR&\leq H(X_{1}^n|S^n_{[2:M]})+ (M-1)H(Y_1^n|S^n_{[2:M]})-H(Y_2^n|S^n_{[2:M]})+nc_6 \\
&=H(X_{1}^n|S^n_{[2:M]})+ (M-2)H(Y_1^n|S^n_{[2:M]}) +H(Y_1^n|S^n_{[2:M]})-H(Y_2^n|S^n_{[2:M]})+nc_6 \\
&\overset{(a)}{\leq}  H(X_{1}^n|S^n_{[2:M]})+ (M-2)H(Y_1^n|S^n_{[2:M]}) +n\cdot o(\log P)+nc_6\\ 
&\overset{(b)}\leq   H(X_{1}^n|X^n_{[2:M]})+ (M-2)H(Y_1^n|S^n_{[2:M]})+n\cdot o(\log P)+nc_6, \label{eqn:cov_3} 
\end{align}
where $(a)$ is from similar steps used to obtain \eqref{eqn:diff} and $(b)$ is due to the Markov chain $X_1^n-(S^n_{[2:M]},\mathbf{h}^n, \mathbf{g}^n)-X^n_{[2:M]}$ that can be shown by marginalizing $p(\mathbf{h}^n,\mathbf{g}^n)p(s_1^n, \cdots, s_M^n)\prod_{t=1}^n \prod_{k=1}^M p(x_{k}(t)|s_k^n, x_k^{t-1}, \mathbf{h}^t)$. 

To bound $H(X_{1}^n|X^n_{[2:M]})$,  we have
\begin{align}
H(Y_1^n)&= H\big(\big\{\sum_{i = 1}^M \lfloor h_i(t) X_i(t)\rfloor\big\}_{t=1}^n \big)\\
&\geq H\big(\big\{\sum_{i = 1}^M \lfloor h_i(t) X_i(t)\rfloor\big\}_{t=1}^n |X_{[2:M]}^n\big)\\
&=  H\big(\big\{ \lfloor h_1(t) X_1(t)\rfloor\big\}_{t=1}^n |X_{[2:M]}^n\big)\\
&=H( X_1^n, \big\{ \lfloor h_1(t) X_1(t)\rfloor\big\}_{t=1}^n  |X_{[2:M]}^n) -H(X_1^n|\big\{ \lfloor h_1(t) X_1(t)\rfloor\big\}_{t=1}^n , X_{[2:M]}^n)\\
&=H( X_1^n  | X_{[2:M]}^n) -H(X_1^n|\big\{ \lfloor h_1(t) X_1(t)\rfloor\big\}_{t=1}^n , X_{[2:M]}^n)\\
&\geq H( X_1^n  | X_{[2:M]}^n)-\sum_{t=1}^nH(X_1(t)|\lfloor h_1(t) X_1(t)\rfloor)\\
&\overset{(a)}{\geq} H( X_1^n| X_{[2:M]}^n)-nc_7, \label{eqn:cov_4}
\end{align}
where $(a)$ is from \cite[Lemma 2]{MukherjeeXieUlukus:arxiv15}. By combining \eqref{eqn:cov_3} and \eqref{eqn:cov_4}, we obtain 
\begin{align}
MnR&\leq   H(Y_1^n)+ (M-2)H(Y_1^n|S^n_{[2:M]})+n\cdot o(\log P)+nc_8\\
&\leq (M-1)H(Y_1^n)+n\cdot o(\log P)+nc_8.
\end{align}
Hence, $d_s\leq \frac{M-1}{M}$. Therefore, for $N=1$, we have 
\begin{align}
d_s\leq \min\left\{\alpha, \frac{M-1}{M}\right\},
\end{align}
which concludes the proof. We note that the proof technique used for the special case of $N=1$ can be extended for the general case of $N\geq 1$, but the resultant bound is loose when $N>1$. \endproof

\section{Conclusion} \label{sec:conclusion}
In this paper, we established the optimal secure d.o.f. of the wiretapped diamond-relay channel where an external eavesdropper not only wiretaps the relay transmissions in the multiple-access part, but also observes a certain number of source-relay links in the broadcast part. The secure d.o.f. was shown to be the same for the cases with and without the knowledge of location of wiretapped links. For the former case, we proposed two constituent jamming schemes. The first scheme is the S-BCJ scheme where the noise symbols are aligned at the destination and the second scheme is the S-CoJ scheme where the noise symbols are beam-formed in the null space of the destination's channel. For the latter case where the location of wiretapped links is unknown, we combined a SNC-like scheme with S-BCJ and S-CoJ in a way that (i) the fictitious message symbols of SNC mask all the message and artificial noise symbols transmitted over the source-relay links and (ii) the fictitious message symbols can be canceled at the destination via beamforming at the relays. We believe that our results on the wiretapped diamond-relay channel are an important  step towards understanding the secrecy capacity of general multi-terminal networks, which remains a largely open problem till date.


\appendices

\section{Proof of the achievability part of Theorem~\ref{thm:less}}\label{appendix:achievability}
Let us first state two theorems that give us achievable secrecy rates for the wiretapped diamond-relay channel. These theorems are obtained by assuming symbol-wise relay operations and then considering the wiretapped diamond-relay channel as the wiretap channel  \cite{CsiszarKorner:78}. These theorems are proved in the end of this appendix. 
\begin{theorem}\label{thm:sec_rate}
For the wiretapped diamond-relay channel with the knowledge of location of wiretapped links, a secrecy rate $R$ is achievable if 
\begin{align}
R\leq I(V;Y_1|\mathbf{h})-I(V;Y_2, S_{T}|\mathbf{h},\mathbf{g})
\end{align}
for some $p(v)p(s_{[1:M]}|v)\prod_{k\in[1:M]}p(x_k|s_k,\mathbf{h})$ such that $H(S_k)\leq C$ and  $\E[X_k^2]\leq P$ for $k\in [1:M]$.  
\end{theorem}
\begin{theorem}\label{thm:sec_rate_no}
For the wiretapped diamond-relay channel without the knowledge of location of wiretapped links, a secrecy rate $R$ is achievable if 
\begin{align}
R\leq I(V;Y_1|\mathbf{h})-I(V;Y_2, S_T|\mathbf{h},\mathbf{g})
\end{align}
for all $T\subseteq [1:M]$ such that $|T|=W$, for some $p(v)p(s_{[1:M]}|v)\prod_{k\in[1:M]}p(x_k|s_k,\mathbf{h})$ such that $H(S_k)\leq C$ and  $\E[X_k^2]\leq P$ for $k\in [1:M]$.  
\end{theorem}

Next, let us present two lemmas used for the proof. The proofs of these lemmas are relegated to the end of this appendix. The first lemma is a result from  Khintchine-Groshev theorem of Diophantine approximation \cite{MotahariOveisGharanMaddahAliKhandani:14}, which plays a key role in the analysis of interference alignment. 
\begin{lem} \label{lemma:Fano}
Consider a random vector $A=(A_1,\cdots,A_{\tau})$ where each of $A_k$'s is a random variable distributed over $\mathcal{C}(\delta,\theta Q)$ for some positive real number $\delta$ and positive integers $\theta$, and $Q$. Assume that a receiver observes $Y$ given as follows: 
\begin{align}
Y=\sum_{k=1}^{\tau}\lambda_k A_k+Z
\end{align}
where $\lambda_1,\cdots, \lambda_{\tau}$ are random variables whose all joint and conditional density functions are bounded\footnote{This condition implies that a space of Lebesque measure zero cannot carry a nonzero probability. } and $Z$ is a Gaussian random variable with zero mean and unit variance. 

Fix arbitrary $\epsilon>0$ and $\gamma>0$. If we choose $Q=P^{\frac{1-\epsilon}{2(\tau+\epsilon)}}$ and $\delta=\frac{\gamma P^{1/2}}{ Q}$, it follows 
\begin{align}
H(A|Y,\lambda_1,\cdots,\lambda_{\tau})\leq o(\log P). 
\end{align}
\end{lem}
The following lemma is used for constructing linear combinations of fictitious message symbols for the case without the knowledge of location of wiretapped links. 
\begin{lem}\label{lemma:mds}
For any positive integers $j$ and $k$ such that $j\leq k$, we can construct a $j\times k$ matrix $\Gamma$ with the following properties: 
\begin{itemize}
\item each element of $\Gamma$ is a non-negative integer smaller than $p$, where $p$ is the smallest prime number greater than or equal to $k$, and
\item any $j$ columns are linearly independent. 
\end{itemize}
\end{lem}

Now, we are ready to prove the achievability part of Theorem~\ref{thm:less}. Note that it is sufficient to show the achievability of  $(\alpha, d_s)=(\frac{M-1}{MN}, \frac{M-1}{M})$ for $N\geq 1$ and $(\alpha, d_s)=(\frac{2}{N}, 1)$ for $N\geq 2$. 

\subsection{Proof of the achievability part of Theorem~\ref{thm:less} for the case with the knowledge of location of wiretapped links} 
 Without loss of generality, let us assume that the first $N$ links are secure, i.e., $T=[N+1:M]$.
\vspace{0.1in}
\subsubsection{S-BCJ scheme achieving  $(\alpha, d_s)=(\frac{M-1}{MN}, \frac{M-1}{M})$}
Let us apply Theorem~\ref{thm:sec_rate} with the following choice of $p(v)p(s_{[1:M]}|v)\prod_{k\in[1:M]}p(x_k|s_k,\mathbf{h})$: 
\begin{align}
V&=(V_{k,j}: k\in [1:N], j\in [1:M-1])\\
S_k&=\begin{cases}
(V_{k,j}: j\in [1:M-1]), & k\in [1:N]\\
\emptyset, &k\in [N+1:M]
\end{cases}\\
X_k&=\begin{cases}
\sum_{j\in [1:M-1]} \mu_{k,j} V_{k,j} + \sum_{j\in [1:N]} \frac{\nu_j}{h_k}U_{k,j}, & k\in [1:N]\\
\sum_{j\in [1:N]} \frac{\nu_j}{h_k}U_{k,j}, & k\in [N+1:M]
\end{cases}
\end{align} 
where $V_{k,j}$'s and $U_{k,j}$'s are independently generated according to $\mbox{Unif}[\mathcal{C}(\delta,Q)]$ for some positive real number $\delta$ and positive integer $Q$ specified later, and $\mu_{k,j}$'s and $\nu_j$'s are independently and uniformly chosen from the interval $[-B, B]$. We note that $H(S_k)\leq C$ and  $\E[X_k^2]\leq P$ for $k\in [1:M]$ are satisfied if 
\begin{align}
(M-1)\log (2Q+1)&\leq C \label{eqn:link}\\
\tilde{\gamma} \delta Q &\leq \sqrt{P}, \label{eqn:power}
\end{align}
where $\tilde{\gamma}=(M-1+NB)B$. Then, the channel outputs are given as  
\begin{align}
Y_1&= \sum_{k=1}^N\sum_{j=1}^{M-1} h_k\mu_{k,j} V_{k,j} + \sum_{k=1}^M \sum_{j=1}^N \nu_j U_{k,j} +Z_1 \\
&=\sum_{k=1}^N\sum_{j=1}^{M-1} h_k\mu_{k,j} V_{k,j} + \sum_{j=1}^N \nu_j\left[\sum_{k=1}^M U_{k,j}\right] +Z_1 \label{eqn:y1_less}\\
Y_2&= \sum_{k=1}^N\sum_{j=1}^{M-1} g_k\mu_{k,j} V_{k,j} + \sum_{k=1}^M \sum_{j=1}^N \frac{g_k\nu_j}{h_k}U_{k,j} +Z_2.
\end{align}
Because $S_T=\emptyset$, Theorem~\ref{thm:sec_rate} says that the following secrecy rate is achievable: 
\begin{align}
R\leq I(V;Y_1|\mathbf{h})-I(V;Y_2|\mathbf{h},\mathbf{g}).\label{eqn:achie}
\end{align}
To derive a lower bound on the RHS of \eqref{eqn:achie}, let us derive a lower and an upper bounds on the first and the second terms in the RHS of \eqref{eqn:achie}, respectively. We will apply Lemma \ref{lemma:Fano} with $\tau \Leftarrow MN$ and hence we choose $Q = P^{\frac{1 - \epsilon}{2(MN + \epsilon)}}$ and $\delta = \frac{ \gamma P^{1/2}}{Q}$ for some $\epsilon>0$ with $\gamma=\tilde{\gamma}^{-1}$ to satisfy the power constraint \eqref{eqn:power}. Now, the first term in the RHS of \eqref{eqn:achie} is bounded as follows: 
\begin{align}
I(V;Y_1|\mathbf{h})&=H(V)-H(V|Y_1,\mathbf{h})\\
&\overset{(a)}\geq \log(2 Q+1)^{N(M-1)} - o(\log{P})\\
&\geq \frac{(1-\epsilon)N(M-1)}{2(MN+\epsilon)}\log P-o(\log P),\label{eqn:same_first}
\end{align}
where $(a)$ is due to Lemma \ref{lemma:Fano} with the substitution of $Y\Leftarrow Y_1$, $\tau\Leftarrow MN$ and $\theta\Leftarrow M$.\footnote{$\theta\Leftarrow M$ is because $\sum_{k=1}^M U_{k,j}$ for each $j\in [1:N]$ is distributed over $\mathcal{C}(\delta, MQ)$ and $\mathcal{C}(\delta,Q)$ is a subset of $\mathcal{C}(\delta, MQ)$.}
Next, we have
\begin{align}
I(V;Y_2|\mathbf{h},\mathbf{g})
&=I(V,U;Y_2|\mathbf{h},\mathbf{g})-I(U;Y_2|V,\mathbf{h},\mathbf{g})\\
&=I(V,U;Y_2|\mathbf{h},\mathbf{g})-H(U)+H(U|Y_{2,\mathrm{eff}},\mathbf{h},\mathbf{g})\\
&=I(V,U;Y_2|\mathbf{h},\mathbf{g})-\frac{(1-\epsilon)MN}{2(MN+\epsilon)}\log P+H(U|Y_{2,\mathrm{eff}},\mathbf{h},\mathbf{g})\\
&\overset{(a)}\leq I(V,U;Y_2|\mathbf{h},\mathbf{g})-\frac{(1-\epsilon)MN}{2(MN+\epsilon)}\log P+o(\log P)\\
&\leq h(Y_2|\mathbf{h},\mathbf{g})-h(Z_2)-\frac{(1-\epsilon)MN}{2(MN+\epsilon)}\log P+o(\log P)\\
&\overset{(b)}\leq \frac{1}{2}\log P -\frac{1}{2}\log 2\pi e -\frac{(1-\epsilon)MN}{2(MN+\epsilon)}\log P+o(\log P)\\
&\leq \frac{\epsilon(MN+1)}{2(MN+\epsilon)}\log P+o(\log P),\label{eqn:same_second}
\end{align}
where $U=(U_{k,j}: k\in [1:M], j\in[1:N])$, $Y_{2,\mathrm{eff}}= \sum_{k=1}^M \sum_{j=1}^N \frac{g_k\nu_j}{h_k}U_{k,j} +Z_2$, $(a)$ follows from Lemma \ref{lemma:Fano} with $Y\Leftarrow Y_{2,\mathrm{eff}}$, $\tau\Leftarrow MN$ and $\theta\Leftarrow 1$, and $(b)$ is because all channel fading coefficients are assumed to be bounded away from zero and infinity. 

By choosing $\epsilon$ sufficiently small, we conclude from  \eqref{eqn:link}, \eqref{eqn:same_first}, and \eqref{eqn:same_second} that $(\alpha, d_s)=(\frac{M-1}{MN}, \frac{M-1}{M})$ is achievable. 

\vspace{0.1in}
\subsubsection{S-CoJ scheme for $N\geq 2$ achieving $(\alpha, d_s)=(\frac{2}{N}, 1)$} 
Let us apply Theorem~\ref{thm:sec_rate} with the following choice of $p(v)p(s_{[1:M]}|v)\prod_{k\in[1:M]}p(x_k|s_k,\mathbf{h})$: 
\begin{align}
V&=(V_{k}: k\in [1:N])\\
S_k&=\begin{cases}
(V_{k}+U_k,U_{[k+1]_N}), & k\in [1:N]\\
\emptyset, &k\in [N+1:M]
\end{cases}\\
X_k&=\begin{cases}
V_k+U_k-\frac{h_{[k+1]_N}}{h_k}U_{[k+1]_{N}}, & k\in [1:N]\\
\emptyset, & k\in [N+1:M]
\end{cases}
\end{align} 
 where $V_{k}$'s and $U_{k}$'s are independently generated according to $\mbox{Unif}[\mathcal{C}(\delta,Q)]$ for some positive real number $\delta$ and positive integer $Q$  specified later.  We note that $H(S_k)\leq C$ and  $\E[X_k^2]\leq P$ for $k\in [1:M]$ are satisfied if 
\begin{align}
2\log (4Q+1)&\leq C \label{eqn:link1}\\
\tilde{\gamma} \delta Q &\leq \sqrt{P},  \label{eqn:power1}
\end{align}
where $\tilde{\gamma}=2+B^2$. 
Then, the channel outputs are given as follows:
\begin{align}
Y_1&= \sum_{k=1}^N  h_k V_k +Z_1 \label{eqn:y1_less4}\\
Y_2&= \sum_{k=1}^N g_k V_k + \sum_{k=1}^N  \left(g_k-\frac{h_k\cdot g_{[k-1]_N}}{h_{[k-1]_N}}\right)U_k +Z_2.
\end{align}

Because $S_T=\emptyset$, from Theorem~\ref{thm:sec_rate}, the following secrecy rate is achievable: 
\begin{align}
R\leq I(V;Y_1|\mathbf{h})-I(V;Y_2|\mathbf{h},\mathbf{g}).\label{eqn:achie1}
\end{align}
To derive a lower bound on the RHS of \eqref{eqn:achie1}, let us derive a lower and an upper bounds on the first and the second terms in the RHS of \eqref{eqn:achie1}, respectively. 
We will apply Lemma \ref{lemma:Fano} with $\tau \Leftarrow N$ and hence we choose $Q = P^{\frac{1 - \epsilon}{2(N + \epsilon)}}$ and $\delta = \frac{ \gamma P^{1/2}}{Q}$ for some $\epsilon>0$ with $\gamma=\tilde{\gamma}^{-1}$ to satisfy the power constraint \eqref{eqn:power1}. Now, the first term in the RHS of \eqref{eqn:achie1} is bounded as follows: 
\begin{align}
I(V;Y_1|\mathbf{h})&=H(V)-H(V|Y_1,\mathbf{h})\\
&\overset{(a)}\geq \log(2 Q+1)^{N} - o(\log{P})\\
&\geq \frac{(1-\epsilon)N}{2(N+\epsilon)}\log P-o(\log P),\label{eqn:same_first1}
\end{align}
where $(a)$ is due to Lemma \ref{lemma:Fano} with the substitution of $Y\Leftarrow Y_1$, $\tau\Leftarrow N$, $\theta\Leftarrow 1$. Next, by applying similar steps used to derive \eqref{eqn:same_second}, we can show 
\begin{align}
I(V;Y_2|\mathbf{h},\mathbf{g})&\leq \frac{\epsilon(N+1)}{2(N+\epsilon)}\log P+o(\log P).\label{eqn:same_second1}
\end{align}

By choosing $\epsilon$ sufficiently small, we conclude from  \eqref{eqn:link1}, \eqref{eqn:same_first1}, and \eqref{eqn:same_second1} that $(\alpha, d_s)=(\frac{2}{N},1)$ is achievable. \endproof


\subsection{Proof of the achievability part of Theorem~\ref{thm:less} for the case without the knowledge of location of wiretapped links} For the case without the knowledge of location of wiretapped links, we superpose a linear combination of fictitious message symbols to each element of $S_k$ in the S-BCJ and S-CoJ schemes.

\vspace{0.1in}

\subsubsection{S-BCJ-SNC scheme achieving $(\alpha, d_s)=(\frac{M-1}{MN}, \frac{M-1}{M})$}
Let $F=(F_k: k\in [1:W(M-1)])$ denote the vector of $W(M-1)$ fictitious message symbols each independently generated according to $\mbox{Unif}[\mathcal{C}(\delta,Q)]$ for some positive real number $\delta$ and positive integer $Q$ to be specified later. To mask each element transmitted over the source-relay links in the S-BCJ scheme, we generate a vector $L=(L_{k,j}: k\in [1:M], j\in [1:M-1])$ of linear combinations of fictitious message symbols by computing $L=F\Gamma$, where $\Gamma$ is an $W(M-1)\times M(M-1)$ matrix that satisfies the properties in Lemma \ref{lemma:mds}, i.e., each element of $\Gamma$ is a non-negative integer smaller than $p$,  where $p$ is the smallest prime number greater than or equal to $M(M-1)$, and every $W(M-1)$ columns of $\Gamma$ are linearly independent. Note that the domain of each element of $L$ is a subset of $\mathcal{C}(\delta, W(M-1)(p-1)Q)$. 
 
Now, we apply Theorem~\ref{thm:sec_rate_no}  with the following choice of $p(v)p(s_{[1:M]}|v)\prod_{k\in[1:M]}p(x_k|s_k,\mathbf{h})$:
\begin{align}
V&=(V_{k,j}: k\in [1:M], j\in [1:M-1])\\ 
S_k&=\begin{cases}
(V_{k,j}+L_{k,j}: j\in [1:M-1]), &k\in [1:N] \\
(L_{k,j}: j\in [1:M-1]), &k\in [N+1:M] \\
\end{cases}
\end{align}
\begin{align}
X_k&=\begin{cases}
\sum_{j\in [1:M-1]} \mu_{k,j} (V_{k,j}+L_{k,j})+ \sum_{j\in [1:N]} \frac{\nu_j}{h_k}U_{k,j}, &k\in [1:N]\\
\sum_{j\in [1:M-1]} \rho_{k,j}L_{k,j} +\sum_{j\in [1:N]} \frac{\nu_j}{h_k}U_{k,j},   & k\in [N+1:M]
\end{cases}
\end{align}
where $V_{k,j}$'s and $U_{k,j}$'s are independently generated according to $\mbox{Unif}[\mathcal{C}(\delta,Q)]$ and $\mu_{k,j}$'s and $\nu_j$'s are independently and uniformly chosen from the interval $[-B, B]$. $\rho_{k,j}$'s are carefully chosen to cancel out the fictitious message symbols at the destination as follows. We first note that because any $W(M-1)$ columns of $\Gamma$ are linearly independent, for $a\in [1:N]$ and $b\in[1:M-1]$, there exists $(\sigma_{a,b|k,j}: k\in [N+1:M], j\in [1:M-1])$ such that 
\begin{align}
L_{a,b}=\sum_{k\in[N+1:M]}\sum_{j\in [1:M-1]}\sigma_{a,b|k,j}L_{k,j}.
\end{align}
To  beam-form each of $F_k$'s in the null space of the destination's channel,  $\rho_{k,j}$ for $k\in [N+1:M]$ and $j\in [1:M-1]$ is chosen as 
\begin{align}
\rho_{k,j}=-\sum_{a\in[1:N]}\sum_{b\in [1:M-1]}\frac{h_a}{h_k}\mu_{a,b}\sigma_{a,b|k,j}. 
\end{align}
Let $\rho_{\max}$ denote the maximum of $\rho_{k,j}$'s. 
We note that $H(S_k)\leq C$ and  $\E[X_k^2]\leq P$ for $k\in [1:M]$ are satisfied if 
\begin{align}
(M-1)\log (2(W(M-1)(p-1)+1)Q+1)\leq C \label{eqn:link2}\\
\tilde{\gamma}\delta Q \leq \sqrt{P},  \label{eqn:power2}
\end{align}
where 
\begin{align}
&\tilde{\gamma}=\max\{(M-1)B(W(M-1)(p-1)+1), (M-1)\rho_{\max}W(M-1)(p-1)\}+NB^2.
\end{align}

Then, the channel outputs are given as 
\begin{align}
Y_1&= \sum_{k=1}^N\sum_{j=1}^{M-1} h_k\mu_{k,j} V_{k,j} + \sum_{k=1}^M \sum_{j=1}^N \nu_j U_{k,j} +Z_1 \\
&=\sum_{k=1}^N\sum_{j=1}^{M-1} h_k\mu_{k,j} V_{k,j} + \sum_{j=1}^N \nu_j\left[\sum_{k=1}^M U_{k,j}\right] +Z_1 \label{eqn:y1_less2} \\
Y_2&= \sum_{k=1}^N\sum_{j=1}^{M-1} g_k\mu_{k,j} V_{k,j} + \sum_{k=1}^M \sum_{j=1}^N \frac{g_k\nu_j}{h_k}U_{k,j}  +\sum_{k=1}^{W(M-1)} \chi_k F_k+Z_2,
\end{align}
where $\chi_k$'s are determined from $\mu_{k,j}$'s, $\rho_{k,j}$'s, $\mathbf{h}$, $\mathbf{g}$, and $\Gamma$. 

From Theorem~\ref{thm:sec_rate_no}, the following secrecy rate is achievable: 
\begin{align}
R\leq  I(V;Y_1|\mathbf{h})-\max I(V;Y_2,S_T|\mathbf{h},\mathbf{g}),\label{eqn:achie_no}
\end{align}
where the maximization is over all $T\subseteq [1:M]$ such that $|T|=W$. To derive a lower bound on the RHS of \eqref{eqn:achie_no}, let us derive a lower and an upper bounds on the first and the second terms in the RHS of \eqref{eqn:achie_no}, respectively.  We will apply Lemma~\ref{lemma:Fano} with $\tau \Leftarrow MN$ and hence we choose $Q = P^{\frac{1 - \epsilon}{2(MN + \epsilon)}}$ and $\delta = \frac{ \gamma P^{1/2}}{Q}$ for some $\epsilon>0$ with $\gamma=\tilde{\gamma}^{-1}$ to satisfy the power constraint \eqref{eqn:power2}. Then, by applying the same bounding techniques used to obtain \eqref{eqn:same_first}, we can obtain
\begin{align}
I(V;Y_1|\mathbf{h})\geq \frac{(1-\epsilon)N(M-1)}{2(MN+\epsilon)}\log P-o(\log P).\label{eqn:same_first2}
\end{align}
Next, let us derive an upper bound on the second term in the RHS of \eqref{eqn:achie_no}. We first fix an arbitrary $T\subseteq [1:M]$ such that $|T|=W$. Note that  
\begin{align}
&I(V;Y_2,S_T|\mathbf{h},\mathbf{g})=I(V;S_T|\mathbf{h},\mathbf{g})+I(V;Y_2|S_T,\mathbf{h},\mathbf{g}). \label{eqn:sec_split}
\end{align}
We first bound the first term in the RHS of \eqref{eqn:sec_split} as follows: 
\begin{align}
I(V;S_T|\mathbf{h},\mathbf{g})&=H(S_T)-H(S_T|V)\\
&\overset{(a)}\leq W(M-1)\log (2(W(M-1)(p-1)+1)Q+1)-H(S_T|V) \\
&= W(M-1)\log (2(W(M-1)(p-1)+1)Q+1)-H(L_{k,j}: k\in T, j\in [M-1]) \\
&\overset{(b)}=W(M-1)\log (2(W(M-1)(p-1)+1)Q+1)-H(F) \\
&=W(M-1)\log (2(W(M-1)(p-1)+1)Q+1) -W(M-1)\log (2Q+1)\\
&=W(M-1)\log \frac{2(W(M-1)(p-1)+1)Q+1}{2Q+1}\\
&=o(\log P), \label{eqn:sp_f}
\end{align}
where $(a)$ is because $S_T$ consists of $W(M-1)$ elements where the domain of each element is a subset of $\mathcal{C}(\delta, (W(M-1)(p-1)+1)Q)$ and $(b)$ is because any $W(M-1)$ columns of $\Gamma$ are linearly independent.
 
Next, the second term in the RHS of \eqref{eqn:sec_split} is bounded as follows: 
\begin{align}
I(V;Y_2|S_T,\mathbf{h},\mathbf{g}) 
&=I(V,U,F;Y_2|S_T,\mathbf{h},\mathbf{g})-I(U,F;Y_2|V,S_T,\mathbf{h},\mathbf{g})\\
&\leq I(V,U,F;Y_2|S_T,\mathbf{h},\mathbf{g})-I(U;Y_2|F,V,S_T,\mathbf{h},\mathbf{g})\\
&\overset{(a)}= I(V,U,F;Y_2|S_T,\mathbf{h},\mathbf{g})-H(U)+H(U|Y_{2,\mathrm{eff}},\mathbf{h},\mathbf{g})\\
&= I(V,U,F;Y_2|S_T,\mathbf{h},\mathbf{g})-\frac{(1-\epsilon)MN}{2(MN+\epsilon)}\log P+H(U|Y_{2,\mathrm{eff}},\mathbf{h},\mathbf{g})\\
&\overset{(b)}\leq I(V,U,F;Y_2|S_T,\mathbf{h},\mathbf{g})-\frac{(1-\epsilon)MN}{2(MN+\epsilon)}\log P+o(\log P)\\
&\leq h(Y_2|\mathbf{h},\mathbf{g})-h(Z_2)-\frac{(1-\epsilon)MN}{2(MN+\epsilon)}\log P+o(\log P)\\
&\leq \frac{1}{2}\log P- \frac{1}{2}\log 2\pi e-\frac{(1-\epsilon)MN}{2(MN+\epsilon)}\log P+o(\log P)\\
&\leq \frac{\epsilon(MN+1)}{2(MN+\epsilon)}\log P+o(\log P)
\end{align}
where  $U=(U_{k,j}: k\in [1:M], j\in[1:N])$, $Y_{2,\mathrm{eff}}= \sum_{k=1}^M \sum_{j=1}^N \frac{g_k\nu_j}{h_k}U_{k,j}+Z_2$, $(a)$ is because $U$ is independent from $F, V, S_T, \mathbf{h}$ and $\mathbf{g}$, and $(b)$ follows from Lemma \ref{lemma:Fano} with $Y\Leftarrow Y_{2,\mathrm{eff}}$, $\tau\Leftarrow MN$ and $\theta\Leftarrow 1$. Since the above bounds do not depend on the choice of $T$, we obtain
\begin{align}
&\max_T I(V;Y_2,S_T|\mathbf{h},\mathbf{g})\leq \frac{\epsilon(MN+1)}{2(MN+\epsilon)}\log P+o(\log P). \label{eqn:less_second2}
\end{align}

By choosing $\epsilon$ sufficiently small,  it follows from \eqref{eqn:link2}, \eqref{eqn:same_first2}, \eqref{eqn:less_second2} that  $(\alpha, d_s)=(\frac{M-1}{MN},\frac{M-1}{M})$ is achievable. 

\vspace{0.1in}
\subsubsection{S-CoJ-SNC scheme achieving $(\alpha, d_s)=(\frac{2}{N}, 1)$}
Let $F=(F_k: k\in [1:2W])$ denote $2W$ fictitious message symbols each independently generated according to $\mbox{Unif}[\mathcal{C}(\delta,Q)]$ for some positive real number $\delta$ and positive integer $Q$ to be specified later. We generate a vector $L=(L_k: k\in [1:2M])$ of  linear combinations of fictitious message symbols by evaluating $L=F\Gamma$, where $\Gamma$ is an $2W\times 2M$ matrix that satisfies the properties in Lemma \ref{lemma:mds}, i.e.,  i.e., each element of $\Gamma$ is a non-negative integer smaller than $p$, where $p$ is the smallest prime number greater than or equal to $2M$, and every $2W$ columns of $\Gamma$ are linearly independent. Note that the domain of each element of $L$ is a subset of $\mathcal{C}(\delta, 2W(p-1)Q)$. 
 
Now, we apply Theorem~\ref{thm:sec_rate_no} with the following choice of $p(v)p(s_{[1:M]}|v)\prod_{k\in[1:M]}p(x_k|s_k,\mathbf{h})$: 
\begin{align}
V&=(V_{k}: k\in [1:N])\\
S_k&=\begin{cases}
(V_{k}+U_k+L_{2k-1},U_{[k+1]_N}+L_{2k}), &k\in [1:N] \\
(L_{2k-1}, L_{2k}), &k\in [N+1:M]
\end{cases}\\
X_k&=\begin{cases}
V_k+U_k+L_{2k-1}-\frac{h_{[k+1]_N}}{h_k}(U_{[k+1]_{N}}+L_{2k}), &k\in [1:N]\\
\rho_{2k-1}L_{2k-1}+\rho_{2k}L_{2k}, &k\in [N+1:M]
\end{cases}
\end{align} 
where $V_{k}$'s and $U_{k}$'s are independently generated according to $\mbox{Unif}[\mathcal{C}(\delta,Q)]$ and $\rho_{k}$'s are chosen to cancel out the fictitious message symbols at the destination as follows. We first note that because any $2W$ columns of $\Gamma$ are linearly independent, for $k\in [1:2N]$, there exists $(\sigma_{k|j}: j\in [2N+1:2M])$ such that 
\begin{align}
L_{k}=\sum_{j\in[2N+1:2M]}\sigma_{k|j}L_{j}.
\end{align}
To  beam-form each of $F_k$'s in the null space of the destination's channel,  $\rho_{2j-1}$ and $\rho_{2j}$ for $j\in [N+1:M]$ are chosen as 
\begin{align}
\rho_{2j-1}&=-\sum_{k\in[1:N]}\frac{h_k}{h_j}\sigma_{2k-1|2j-1}-\sum_{k\in[1:N]}\frac{h_{[k+1]_N}}{h_j}\sigma_{2k|2j-1}\\
\rho_{2j}&=-\sum_{k\in[1:N]}\frac{h_k}{h_j}\sigma_{2k-1|2j} -\sum_{k\in[1:N]}\frac{h_{[k+1]_N}}{h_j}\sigma_{2k|2j}.
\end{align}
Let $\rho_{\max}$ denote the maximum of $\rho_k$'s. We note that $H(S_k)\leq C$ and  $\E[X_k^2]\leq P$ for $k\in [1:M]$ are satisfied if 
\begin{align}
2\log (4(W(p-1)+1)Q+1)\leq C \label{eqn:link3}\\
\tilde{\gamma} \delta Q \leq \sqrt{P},  \label{eqn:power3}
\end{align}
where
\begin{align}
\tilde{\gamma}&=\max\{2+2W(p-1)+B^2(1+2W(p-1)),4\rho_{\max}W(p-1)\}.
\end{align}

Then, the channel outputs are given as follows:
\begin{align}
Y_1&= \sum_{k=1}^N  h_k V_k +Z_1 \label{eqn:y1_less3}\\
Y_2&= \sum_{k=1}^N g_k V_k + \sum_{k=1}^N \left(g_k-\frac{h_k\cdot g_{[k-1]_N}}{h_{[k-1]_N}}\right)U_k  +\sum_{k=1}^{2W} \chi_k F_k+Z_2,
\end{align}
where $\chi_k$'s are determined from $\rho_{k}$'s, $\mathbf{h}$, $\mathbf{g}$, and $\Gamma$. 

From Theorem~\ref{thm:sec_rate_no}, the following secrecy rate is achievable: 
\begin{align}
R\leq  I(V;Y_1|\mathbf{h})-\max I(V;Y_2,S_T|\mathbf{h},\mathbf{g}),\label{eqn:achie_no2}
\end{align}
where the maximization is over all $T\subseteq [1:M]$ such that $|T|=W$. As in the S-CoJ scheme, we will apply Lemma~\ref{lemma:Fano} with $\tau \Leftarrow N$ and hence we choose $Q = P^{\frac{1 - \epsilon}{2(N + \epsilon)}}$ and $\delta = \frac{ \gamma P^{1/2}}{Q}$ for some $\epsilon>0$ with $\gamma=\tilde{\gamma}^{-1}$ to satisfy the power constraint \eqref{eqn:power3}. Now, by applying the same bounding techniques used to obtain \eqref{eqn:same_first1}, the first term in the RHS of \eqref{eqn:achie_no2} is bounded as follows: 
\begin{align}
I(V;Y_1|\mathbf{h})&\geq \frac{(1-\epsilon)N}{2(N+\epsilon)}\log P-o(\log P).\label{eqn:same_first3}
\end{align}

Next, let us derive an upper bound on the second term in the RHS of \eqref{eqn:achie_no2}. We first fix an arbitrary $T\subseteq [1:M]$ such that $|T|=W$. Note that  
\begin{align}
I(V;Y_2,S_T|\mathbf{h},\mathbf{g})
&=I(V;S_T|\mathbf{h},\mathbf{g})+I(V,U,F;Y_2|S_T,\mathbf{h},\mathbf{g})-I(U,F;Y_2|V,S_T,\mathbf{h},\mathbf{g}),\label{eqn:sec_split2}
\end{align}
where $U=(U_{k}: k\in [1:N])$. Then, by applying similar bounding techniques used to obtain \eqref{eqn:sp_f}, the first term in the RHS of \eqref{eqn:sec_split2} can be bounded as follows: 
\begin{align}
I(V;S_T|\mathbf{h},\mathbf{g})\leq o(\log P). \label{eqn:f1}
\end{align} 
Next, the second  term in the RHS of \eqref{eqn:sec_split2} is bounded as follows: 
\begin{align}
I(V,U,F;Y_2|S_T,\mathbf{h},\mathbf{g})&\leq h(Y_2|\mathbf{h},\mathbf{g})-h(Z_2)\\
&\leq \frac{1}{2} \log P + o(\log P). \label{eqn:f2}
\end{align}
Finally, the third term in the RHS of \eqref{eqn:sec_split2} is bounded as follows: 
\begin{align}
I(U,F;Y_2|V,S_T,\mathbf{h},\mathbf{g})&=H(U,F|V,S_T)-H(U,F|Y_2,V,S_T,\mathbf{h},\mathbf{g})\\
&\overset{(a)}=H(U,F,V)-H(V,S_T)-H(U,F|Y_2,V,S_T,\mathbf{h},\mathbf{g})\\
&\overset{(b)}\geq H(U)+H(F)-H(S_T) -H(U,F|Y_2,V,S_T,\mathbf{h},\mathbf{g})\\
&\overset{(c)}\geq H(U)-H(U,F|Y_2,V,S_T,\mathbf{h},\mathbf{g})-o(\log P)\\
&= H(U)\!-H(U|Y_{2},V,S_T,\mathbf{h},\mathbf{g}) \!-H(F|U,Y_{2},V,S_T,\mathbf{h},\mathbf{g})\!-o(\log P)\label{eqn:eav_split2}
\end{align}
where $(a)$ is because $S_T$ is a function of $U, F$, and $V$, $(b)$ is because $U, F, V$ are mutually independent, and $(c)$ is by applying similar steps used to obtain \eqref{eqn:sp_f}.

To bound $H(U|Y_{2},V,S_T,\mathbf{h},\mathbf{g})$, we note that $S_T$ can be represented as a vector of length $2W$ given as 
\begin{align}
S_T=V\Lambda_{V,T}+U\Lambda_{U,T}+F\Gamma_{T}
\end{align}
where $\Lambda_{V,T}$ and $\Lambda_{U,T}$ are $N\times 2W$  matrices that are determined from our choice of $S_k$'s and $\Gamma_{T}$ is a $2W\times 2W$ submatrix of $\Gamma$ corresponding to the choice of $T$. Because any $2W$ columns of $\Gamma$ are linearly independent, the inverse matrix $\Gamma_T^{-1}$ of $\Gamma_T$ exists and hence we can represent $F$ as follows: 
\begin{align}
F=(S_T-V\Lambda_{V,T}-U\Lambda_{V,T})\Gamma_T^{-1}. \label{eqn:F}
\end{align}
Now, by substituting \eqref{eqn:F} for $F$ in $Y_2$, $Y_2$ can be represented as a function of $V, U, S_T$ and $Z_2$. Hence, the effective channel output $Y_{2,\mathrm{eff}}$, where the contribution from $V$ and $S_T$ are canceled out, is a linear combination of $U_k$'s and $Z_2$. By applying Lemma \ref{lemma:Fano} with $Y\Leftarrow Y_{2,\mathrm{eff}}$, $\tau\Leftarrow N$, and $\theta\Leftarrow 1$, it follows that 
\begin{align}
H(U|Y_{2},V,S_T,\mathbf{h},\mathbf{g})&=H(U|Y_{2,\mathrm{eff}},\mathbf{h},\mathbf{g})\\
&\leq o(\log P). \label{eqn:eav_split3}
\end{align}
Furthermore, due to \eqref{eqn:F}, it follows that 
\begin{align}
H(F|U,Y_{2},V,S_T,\mathbf{h},\mathbf{g})=0\label{eqn:eav_split4}
\end{align}
Therefore, we have
\begin{align}
I(U,F;Y_2|V,S_T,\mathbf{h},\mathbf{g})
&\geq H(U)-o(\log P)\\
&=\frac{(1-\epsilon)N}{2(N+\epsilon)}\log P-o(\log P). \label{eqn:f3}
\end{align}
From \eqref{eqn:f1}, \eqref{eqn:f2}, \eqref{eqn:f3}, we obtain 
\begin{align}
I(V;Y_2,S_T|\mathbf{h},\mathbf{g})&\leq \frac{\epsilon(N+1)}{2(N+\epsilon)}\log P+o(\log P). \label{eqn:less_second3}
\end{align}

By choosing $\epsilon$ sufficiently small, it follows from \eqref{eqn:link3}, \eqref{eqn:same_first3}, \eqref{eqn:less_second3} that  $(\alpha, d_s)=(\frac{2}{N},1)$ is achievable. 
\endproof

\subsubsection*{Proof of Theorems \ref{thm:sec_rate} and \ref{thm:sec_rate_no}}
Let us restrict relay operations to be symbol-wise, i.e., at time $t\in [1:n]$, relay $k\in[1:M]$ (randomly) maps $(S_k(t), \mathbf{h}(t))$ to $X_k(t)$ according to $p(x_k|s_k, \mathbf{h})$. Then, for the case with the knowledge of location of wiretapped links,  the wiretapped diamond-relay channel can be considered as the wiretap channel with channel input $(S_k: k\in [1:M])$, legitimate channel output $(Y_1,\mathbf{h})$, eavesdropper channel output $(Y_2, S_T,\mathbf{h},\mathbf{g})$, and channel distribution marginalized from $p(\mathbf{h},\mathbf{g})\prod_{k\in[1:M]}p(x_k|s_k,\mathbf{h})p(y_1,y_2|x_{[1:M]},\mathbf{h},\mathbf{g})$.  Then, Theorem~\ref{thm:sec_rate} is immediate from \cite{CsiszarKorner:78}. For the case without the knowledge of location of wiretapped links, Theorem~\ref{thm:sec_rate_no} can be proved in a similar manner by assuming there are multiple eavesdroppers each of which observes every different $S_T$ such that $|T|=W$. \endproof

\subsubsection*{Proof of Lemma \ref{lemma:Fano}}
According to the Khintchine-Groshev theorem of Diophantine approximation \cite{MotahariOveisGharanMaddahAliKhandani:14}, for any $\epsilon>0$ and almost all  $\lambda_1,\cdots, \lambda_{\tau}$  except a set of Lebesque measure zero, there exists a constant $k_{\epsilon}$ such that 
\begin{align}
|\sum_{k=1}^{\tau}\lambda_k q_k|>\frac{k_{\epsilon}}{\max_{k}|q_k|^{\tau-1+\epsilon}}
\end{align}
holds for all $(q_1,\cdots, q_{\tau}) \neq \mathbf{0}\in \mathbbm{Z}^{\tau}$. 
Hence, the minimum distance $d_{\min}(\lambda_1,\cdots,\lambda_{\tau})$ between the points in $\{\sum_{k=1}^{\tau}\lambda_k a_k: a_k\in \mathcal{C}(\delta,\theta Q)\}$ is bounded as follows: 
\begin{align}
d_{\min}(\lambda_1,\cdots,\lambda_{\tau}) \geq \frac{\delta k_{\epsilon}}{(\theta Q)^{\tau-1+ \epsilon}} \label{eqn:d_min}
\end{align}
for almost all  $\lambda_1,\cdots, \lambda_{\tau}$ except a set of Lebesque measure zero. 
 
Now, when $(\lambda_1,\cdots,\lambda_{\tau})$ is known to the receiver, let $\hat{A}(\lambda_1,\cdots,\lambda_{\tau})$ denote the estimate of $A$ which is chosen as follows:
\begin{align}
\hat{A}(\lambda_1,\cdots,\lambda_{\tau})=\argmin_{ (a_1,\cdots, a_{\tau}): \atop a_k\in \mathcal{C}(\delta,\theta Q)} \big|Y-\sum_{k=1}^{\tau}\lambda_k a_k\big|.
\end{align}
Then, for the choice of $Q=P^{\frac{1-\epsilon}{2(\tau+\epsilon)}}$ and $\delta=\frac{\gamma P^{1/2}}{Q}$, we have 
 \begin{align}
P(\hat{A}(\lambda_1,\cdots,\lambda_{\tau}) \neq A ) &\leq \E\Big[\exp\left(-\frac{d_{\min}^2(\lambda_1,\cdots,\lambda_{\tau})}{8}\right)\Big]\\
    &\overset{(a)}\leq \exp\left(-\frac{\delta^2 k_{\epsilon}^2}{8(\theta Q)^{2(\tau-1 + \epsilon) }}\right)\\
 &\leq \exp\left(-\frac{\gamma^2 k_{\epsilon}^2P^{\epsilon}}{8\theta ^{2(\tau-1+\epsilon)}}\right),
 \end{align}
 where $(a)$ is because the probability that $\lambda_1,\cdots, \lambda_{\tau}$ are included in a set of Lebesque measure zero is zero by assumption. 

 According to the Fano's inequality, it follows that
\begin{align}
H(A|Y,\lambda_1,\cdots,\lambda_{\tau})
 &\leq  H(A|\hat{A}(\lambda_1,\cdots,\lambda_{\tau}))\\
&\leq 1 + P(\hat{A}(\lambda_1,\cdots,\lambda_{\tau})\neq A) \log(|A| - 1)\\
& \leq 1 +  \exp\left(-\frac{\gamma^2k_{\epsilon}^2P^{\epsilon}}{8\theta^{2(\tau-1+\epsilon)}}\right)\log{(2\theta Q )^{\tau}}\\
& = o(\log{P}),
\end{align}
which concludes the proof. \endproof

\subsubsection*{Proof of Lemma \ref{lemma:mds}}
Let us show that  the generator matrix $\Gamma$ of a Reed-Solomon code \cite{ReedSolomon:60} with block length $k$, message length $j$, and alphabet size $p$ satisfies the aforementioned properties. First, because $\Gamma$ is over the prime field $\mbox{GF}(p)$, its element is an integer in the range $0, \cdots, p-1$, and hence the first property is satisfied.  For the second property, we first note that the sum, the difference and the product over $\mbox{GF}(p)$ are computed by taking the modulo $p$ of the integer result. Since a Reed-Solomon code is an MDS-code, any $j$ columns of $\Gamma$ are linearly independent over $\mbox{GF}(p)$, which means that the determinant of any $j\times j$ submatrix of $\Gamma$ \emph{evaluated  over the prime field $\mbox{GF}(p)$} is nonzero. This implies that the determinant of any $j\times j$ submatrix of $\Gamma$  \emph{ evaluated over real numbers} is nonzero, and hence the second property is satisfied. \endproof


\end{document}